\documentclass[journal]{IEEEtran}

\usepackage{graphics,epsf,psfrag,amsmath,amssymb,amsfonts,verbatim,cite}
\usepackage[mathscr]{eucal}

\def\ACSP{/research/ACSP}

\def\boxit#1{\vbox{\hrule\hbox{\vrule\kern3pt
        \vbox{\kern3pt#1\kern3pt}\kern3pt\vrule}\hrule}}

\def\reals{ { {\rm  I \kern-0.15em R }  } }
\def\complex{ {\,{{\rm C} \kern-0.50em \raise0.20ex {  |}}\, }}

\def\Rbf{{\bf R}}

\def\Gc{{\cal G}}

\def\defeq{{\stackrel{\Delta}{=}}}
\def\Rxx{\Rbf_{\ssstyle X\kern-.1em X}}

\let\ssstyle=\scriptscriptstyle

\def\eg{{\it e.g.,}}

\def\ie{{\it i.e.,\ \/}}
\def\Kout{\setbox1=\hbox{\Huge\bf K}\hbox to
1.05\wd1{\hspace{.05\wd1}
}}
\def\Sout{\setbox1=\hbox{\Huge\bf S}\hbox to 1.05\wd1{\hspace{.05\wd1}
}}

\def\scalefig#1{\epsfxsize #1\textwidth}

\newtheorem{lemma}{Lemma}

\newcommand{\Amsc}{\mathscr{A}}
\newcommand{\Bmsc}{\mathscr{B}}
\newcommand{\Cmsc}{\mathscr{C}}

\newcommand{\mbbE}{\mathbb{E}}

\newcommand{\beq}{\begin{equation}}
\newcommand{\eeq}{\end{equation}}

\title{Capacity of Cooperative Fusion in the Presence of\\ Byzantine Sensors}

\author{Oliver~Kosut and~Lang~Tong,~\IEEEmembership{Fellow,~IEEE}
\thanks{This work is supported in part by TRUST
(The Team for Research in Ubiquitous Secure Technology), which
receives support from the National Science Foundation (NSF award
number CCF-0424422) and the following organizations: Cisco,
ESCHER, HP, IBM, Intel, Microsoft, ORNL, Qualcomm, Pirelli, Sun
and Symantec, and the U. S. Army Research Laboratory under the
Collaborative Technology Alliance Program, Cooperative Agreement
DAAD19-01-2-0011. The U. S. Government is authorized to reproduce
and distribute reprints for Government purposes notwithstanding
any copyright notation thereon. A version of this paper was submitted to the   	
44th Annual Allerton Conference on Communication, Control, and Computing.}}

\begin{document}
\maketitle

\begin{abstract}
The problem of cooperative fusion in the presence of Byzantine sensors is considered. An information theoretic formulation is used to characterize the Shannon capacity of sensor fusion. It is shown that when less than half of the sensors are Byzantine, the effect of Byzantine attack can be entirely mitigated, and the fusion capacity is identical to that when all sensors are honest. But when at least half of the sensors are Byzantine, they can completely defeat the sensor fusion so that no information can be transmitted reliably. A capacity achieving transmit-then-verify strategy is proposed for the case that less than half of the sensors are Byzantine, and its error probability and coding rate is analyzed by using a Markov decision process modeling of the transmission protocol.
\end{abstract}


\section{Introduction}
\PARstart{W}{ireless} sensor networks are not physically secure;
they are vulnerable to various attacks.  For example,
sensors may be captured and analyzed such that
the attacker gains inside information about
the communication scheme and networking protocols. The attacker can then reprogram the compromised
sensors and use them to launch the so-called
Byzantine attack.  This paper presents an information theoretic
 approach to sensor fusion in the presence of
 Byzantine sensors.

\subsection{Cooperative Sensor Fusion}
\begin{figure}[htb]
\centerline{
\begin{psfrags}
\psfrag{q}[c]{$q(y|x)$}
\psfrag{Fusion Center}[c]{\small Fusion center}
\psfrag{Byzantine Sensor}[l]{\small Byzantine sensor}
\psfrag{Honest Sensor}[l]{\small Honest sensor}
\scalefig{0.4}\epsfbox{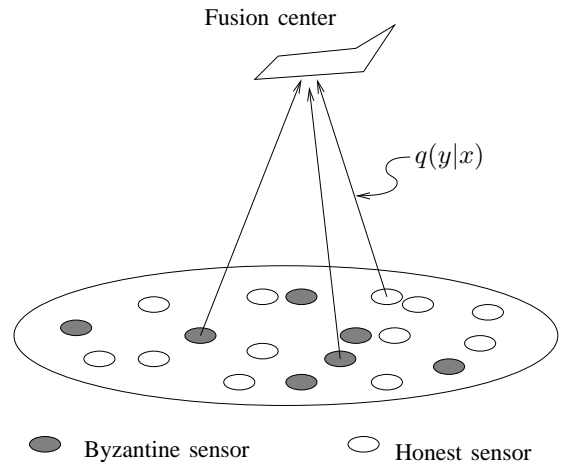}
\end{psfrags}}
\caption{Cooperative sensor fusion in the presence of Byzantine sensors.}
\label{fig:fusion}
\end{figure}

We consider the problem of cooperative sensor fusion
as illustrated in Fig.~\ref{fig:fusion} where the fusion
center extracts information from a sensor field.  By cooperative
fusion we mean that sensors first reach a consensus among themselves
about the fusion message. They then deliver the agreed message
to the fusion center collaboratively.  We will not be
concerned with how sensors reach consensus in this paper; see \eg \cite{Barborak&Malek&Dahbura:93ACM}.
We focus instead on achieving the maximum rate of sensor fusion.

The sensor fusion problem is trivial if the consensus is perfect, \ie all the sensors agree on the same fusion message.
If the fusion center can only communicate with one sensor at a time,
and  there is no limit on how many times a sensor can transmit
(\ie no energy constraints),
there is no difference between having a single sensor delivering
the message and having any number of sensors transmitting the message collaboratively.
The capacity of such an
ideal fusion is given by the classical Shannon theory
\begin{equation} \label{eq:c0}
C = \max_{p(x)} I(X;Y)
\end{equation}
where $X$ is the transmitted symbol by a sensor, $Y$ the received symbol, and $p(x)$
the distribution used to generate the codebook.
Also for this case, even if there is a feedback channel from the
fusion center to  sensors, the capacity does not increase \cite{Cover&Thomas:book}.

Cooperative fusion becomes important if consensus
cannot be reached, \ie there is a probability $\beta>0$ that
a particular sensor is misinformed about what message
to transmit. Thus there is a positive probability that
a particular sensor communicating with the fusion center is
delivering the wrong message. It is no longer obvious what
the capacity of sensor fusion is.  In \cite{Yang&Tong:05IT}, a number
of sensor fusion models are considered, and the fusion capacity
is obtained for several cases.  Most relevant to
this paper is the fusion model in which there is a feedback channel from the fusion
center to individual sensors, and the  fusion center
polls specific sensors for transmissions.  Optimized among all
polling strategies, it is shown that, for any $\beta<1$,
the fusion capacity is also given by $C$ in (\ref{eq:c0}).
The strategy given in \cite{Yang&Tong:05IT} can be characterized
as ``identify-then-transmit'' by first using an
asymptotically negligible number of transmissions to identify
a sensor that is correctly informed then letting that sensor transmit the
entire codeword.

\subsection{Byzantine Attack and Related Work}
The problem considered in this paper is when a fraction $\beta$ of
sensors are Byzantine sensors. The goal of these Byzantine
sensors is to disrupt the sensor fusion collaboratively.

We assume that Byzantine sensors have full knowledge
of the system and impose no restriction on
what they can transmit.  In particular, Byzantine sensors know
the transmission strategy including the codebook and the
polling strategy of the fusion center. They also know, of course,
the correct fusion message.  Unlike the misinformed
sensors that transmit
randomly selected messages  in \cite{Yang&Tong:05IT}, Byzantine
sensors can be malicious sometimes and behave in other times
as honest sensors in order to evade detection
by the fusion center.  Furthermore,  they can
coordinate among themselves (unknown to both the honest sensors
and the fusion center) to launch the so-called Byzantine
attack.  As a result, the capacity achieving coding and transmission
strategies developed in \cite{Yang&Tong:05IT} are no longer applicable.

The notion of Byzantine attack has its root in the Byzantine generals problem \cite{Lamport&Shostak&Pease:82ACM,Dolev:82} in which
a clique of traitorous generals conspire to prevent loyal generals to form consensus.
It was shown in \cite{Lamport&Shostak&Pease:82ACM} that
 consensus in the presence of Byzantine attack is possible if and only if less than
 $\frac{1}{3}$ of the generals
are traitorous. Relaxing the strict definition of consensus of
\cite{Lamport&Shostak&Pease:82ACM},
Pfitzmann and Waidner uses an information theoretic approach to show that
Byzantine general problem can be solved for an arbitrarily large fraction
of Byzantine nodes \cite{Pfitzmann&Waidner:96RPT}.  These and other Byzantine
consensus results \cite{Barborak&Malek&Dahbura:93ACM} are
relevant to the current paper only in that they deal with the consensus process
prior to sensor fusion.

Countering Byzantine attacks in communication networks has
also been studied in the past by many authors.  See the earlier
work of Perlman \cite{Perlman:88thesis} and also
more recent review \cite{Zhou&Haas:99,Hu&Perrig:04}.
An information theoretic network coding approach to
Byzantine attack is presented in \cite{Ho&etal:04ISIT}.
Karlof and Wagner \cite{Karlof&Wagner:03SNPA}
consider routing
security in wireless sensor networks. They introduce different
kinds of attacks  and analyze
security risks of all major existing sensor network routing protocols. Countermeasures and design
considerations for secure routing in sensor networks are also discussed. It is shown that cryptography alone is not enough; careful protocol design is necessary.

There has been limited attempt in dealing
with Byzantine attacks for sensor fusion.  The problem of optimal
Byzantine attack of sensor fusion for distributed detection
is considered in \cite{Marano&Matta&Tong:06Asilomar} where the authors
show that exponentially decaying detection error probabilities
can still be maintained if and only if
the fraction of Byzantine sensors is less than half.
A witness-based approach to sensor fusion is proposed by Du et. al.
\cite{Du&Deng&Han&Varshney:03GCom} where the fusion center and a set of
witnesses jointly authenticate the fusion data by the use of the Message Authentication Code.
The authors of \cite{Du&Deng&Han&Varshney:03GCom} are concerned with the trustworthiness of the fusion center. In contrast, we
address the problem of sensor fusion with malicious sensors attacking the fusion center from within.

\subsection{Main Result and Organization}

The main result of this paper is to show that, if polling of the
fusion center is allowed, and the polling is perfect, the capacity of
sensor fusion in the presence of Byzantine attack is again
$C$ in (\ref{eq:c0}) when $\beta<\frac{1}{2}$ and 0 when $\beta\ge\frac{1}{2}$.

The converse of the result holds trivially for $\beta<\frac{1}{2}$ because the capacity
of the sensor fusion in the absence of Byzantine sensors
is $C$. For $\beta\ge\frac{1}{2}$, we show that it is possible for the Byzantine sensors 
to completely defeat the fusion center and honest sensors by setting things up so that 
exactly half the sensors act honestly with the true message and the other half also act 
honestly but with a false message. It is thus impossible for the fusion center to 
distinguish the set transmitting the true message from the set transmitting the false 
one, so it cannot decode the true message with probability more than $\frac{1}{2}$.

  To show the achievability for $\beta<\frac{1}{2}$, we propose a
 transmission and coding strategy different
from that for misinformed sensors \cite{Yang&Tong:05IT}, for which the
capacity achieving strategy can be called ``identify-then-transmit'',
where the fusion center first identifies an honest sensor, then receives the entire message from that sensor.
Here we must deal with the situation in which a Byzantine sensor may pretend to be
an honest sensor.  The key idea is one of ``transmit-then-verify''.
Specifically, we first commit a sensor (Byzantine or honest)
to transmit part of a codeword and then verify if the sensor is
trustworthy.  After a sensor has transmitted, the fusion center
verifies the transmission using  a random binning procedure.
Under this procedure, a Byzantine sensor
either has to act honestly or reveal with high probability its identity.
We then have to show that the overhead in the
verification diminishes as the length of the codeword increases.

This paper is organized as follows.  In Section~\ref{sec:model}, we
present models for sensors, communication channels, and network setup.
The main result is given and sketch of proofs are presented in Section~\ref{sec:main}.
We conclude in Section~\ref{sec:conc}.

\section{Model and Definitions} \label{sec:model}
\subsection{Fusion Network and Communication Channels}
A sensor is \emph{Byzantine} if it can behave arbitrarily.  A sensor
is \emph{honest} if it behaves only according to the specified protocol.
Let $\beta$ be the probability that a randomly selected sensor is Byzantine.
With probably $1-\beta$, a randomly chosen sensor is honest.
We assume that the sensor network is large in
the sense that there are an infinite number of elements.
This assumption ensures that the probability of all nodes being Byzantine is zero.

Sensors can communicate with the fusion center directly, and the transmissions are time slotted.
We assume that the uplink channel from each sensor to the fusion center is
a Discrete Memoryless Channel (DMC) $\{\mathcal{X},\mathcal{Y},q(y|x)\}$ where $\mathcal{X}$ is the input alphabet,
$\mathcal{Y}$ the output alphabet, and $q(y|x)$ the transition probability of the channel.
The assumption of identical channel is restrictive and synchronization difficult
when the network is large and the fusion center stationary. The assumed model is reasonable, however,
if the fusion center is a mobile access point that can travel around the network, and a sensor
only transmits to the fusion center when it is activated by and synchronized to
the fusion center.

We assume that there is a polling channel from the fusion center to each sensor.
Since the fusion center is not power limited, we assume the polling channel is
error free with infinite capacity.

\subsection{Transmission Protocol}

Before sensor fusion starts, we assume that the sensor network, without error,
has agreed upon a fusion message $W\in\{1,\cdots,M\}$ that is uniformly distributed. 
The code is in general variable length and dynamically generated, so there is no single fixed 
codebook. However, we assume that the sensors may have any number of fixed codebooks to use 
as pieces of the code.

The fusion center polls one node to transmit one symbol at each time slot.
At time $t$, the fusion center polls node $K_t$ to transmit a symbol $X_t$.
The symbol received by the fusion center is then $Y_t$.
The fusion center may choose $K_t$ based on previously received symbols $Y^{t-1}$
and polling history $K^{t-1}$. Since the polling channel has infinite capacity, $K_t$
may choose $X_t$ based on all symbols previously received by the fusion center $Y^{t-1}$,
the polling history $K^{t-1}$, and anything else the fusion center chooses to send to it.
It may also base $X_t$ on all previous transmissions that it has made itself,
but not those made by other sensors, and of course the message $W$.

If a sensor is Byzantine, it may also base its choice of $X_t$
on all transmitted symbols, including those sent by honest sensors,
and any additional information the fusion center sends to any sensor.
We also assume that the Byzantine sensors know the algorithm the fusion
center and honest sensors are using, and that they may communicate securely
among themselves with zero error.

After the fusion center receives $Y_t$, it decides whether  to
continue polling based on $Y^t$ and $K^t$. If it decides to continue, then it moves on to the
next time slot $t+1$ and starts the polling step again.  Otherwise, it decodes
based on collected observations.

\subsection{Achievable Rates and Capacity}

Let $N$ be the random variable representing the total number of
symbols sent in a coding session. Once the fusion center decides it is done polling,
it decodes the global
message based on $Y^N$ and $K^N$. The decoded message is denoted by
$\hat{W}\in\{1,\cdots,M\}$. A decoding error occurs if $\hat{W}\ne W$.

The \emph{rate} of a code is defined as
\[R\triangleq \frac{\log(M)}{\mbbE(N)},\]
where $M$ is the number of messages and $\mbbE(N)$ is the expected number
of symbols transmitted during a coding session. The probability of error is defined as $P_e\triangleq\Pr(\hat{W}\ne W)$, where $W$ is the message, uniformly selected from $\{1,\cdots,M\}$, and $\hat{W}$ is the decoded message. $P_e$ will in general depend on the actions of the Byzantine sensors. A rate $R$ is called \emph{achievable} if for any given error $\epsilon>0$ and any choice of actions by the Byzantine sensors, there exists a code with rate larger than $R-\epsilon$ and probability of error less than $\epsilon$. The \emph{capacity} of this system is defined as the maximum of all achievable rates.

\section{Fusion Capacity}\label{sec:main}

The main result of this paper is given by the following theorem that characterizes
 the capacity for the fusion network described in Sec~\ref{sec:model}.

\emph{Theorem:} The capacity of this system is
\[C^{\mathrm{byz}}=\left\{
\begin{array}{ll}
C,&\textrm{if }\beta<1/2\\
0,&\textrm{if }\beta\ge1/2
\end{array}\right.\]
where $C$ is defined in (\ref{eq:c0}).

A sketch of the proof of this theorem follows. In Subsection~\ref{sub:converse}, we prove the converse. In Subsection~\ref{sub:strategy}, we describe the coding strategy used to prove achievability. In Subsection~\ref{sub:events}, we define some error events and discuss the error probability. Finally, in Subsection~\ref{sub:rate}, we discuss the rate of this coding scheme.

\subsection{Converse}\label{sub:converse}

Suppose that $\beta=0$ and that all the sensors may communicate with each other with zero error. Certainly these assumptions cannot decrease the capacity for any $\beta$. Since the sensors can communicate with each other, we can think of the entire sensor network as a single encoder for the DMC with perfect feedback, since the sensors are allowed to know all previously received symbols by the fusion center. Thus under these assumptions this system reduces to a point-to-point DMC with perfect feedback. In that system, the feedback does not increase capacity \cite{Cover&Thomas:book}, so the capacity is $C$. Thus, the capacity of the sensor network with Byzantine sensors cannot have capacity greater than this, so $C^{\mathrm{byz}}\le C$ for all $\beta$.

Next we show that if $\beta\ge\frac{1}{2}$, then $C^{\mathrm{byz}}=0$. To do this, we will show that for any algorithm to be used by the fusion center and honest sensors, the Byzantine sensors will be able to make it impossible for the probability of error to be made arbitrarily small. The scheme performed by the Byzantine sensors to accomplish this is as follows. They divide themselves into two groups, one with $\frac{1}{2}$ of the sensors, and one with $\beta-\frac{1}{2}$ of the sensors. The sensors in the latter group act exactly like honest sensors. Since there is no way for the honest sensors to know anything that the Byzantine sensors do not, it will be impossible to distinguish an honest sensor from a Byzantine sensor acting honestly. The sensors in the former group also act exactly like honest sensors, but with a message different from the true one. Thus exactly half of the sensors---the honest sensors plus the Byzantine sensors that act honestly---will act honestly with the true message. The other half of the sensors---the rest of the Byzantine sensors---also act honestly but with an incorrect message. Thus, since the number of sensors in these two groups is the same, no matter what the fusion center does, it will not be able to determine which half is reporting the true message and which half is reporting the false one, so it will not be able to decode the true message with probability greater than $\frac{1}{2}$. Therefore the converse of the theorem holds.

\subsection{Coding Strategy}\label{sub:strategy}
To prove the direct part of the theorem, we first describe the
coding strategy that will achieve this rate. The coding scheme can be described
as a ``transmit-then-verify'' procedure. In other words, first we ask a sensor to send part of
the message to the fusion center. After that, the fusion center polls other sensors
to verify whether the received information is correct. Thus, if a Byzantine sensor is
selected to transmit the message, it can send erroneous information, but then with high
probability it will be discovered to be erroneous in the ``verify'' step. The Byzantine sensor
can send the true information, but then it will be verified, so the fusion center now has that information,
and knows it to be correct. As long as the fusion center always verifies any information it receives,
the Byzantine sensors can never get any false information through. The best they can do is to prolong
the coding process, but we will show that this additional overhead can be made to be negligible.

The coding strategy is as follows. We first break the message up into $v$ \emph{chunks},
such that each chunk contains an equal part of the information in the message, and the message will
be perfectly reconstructible given all the chunks. These chunks could be, for example, the $v$ digits representing the message
$W$ when it is written as a number in a particular base.  The fusion center will try to obtain
the $v$ chunks one at time, and verify that each chunk obtained is from an honest transmission.

Next we describe the two codebooks to be used in the uplink transmission over the DMC $q(y|x)$.
Take any $\epsilon>0$ and $R<C$. Let the number of possible messages $M=2^{nR}$,
so that the message set is $\{1,\cdots,2^{nR}\}$ and the set of all possible chunks is $\{1,\cdots,2^{nR/v}\}$.
The first codebook $\Gc_1$ is a $(2^{nR/v},n/v,\epsilon)$ code to transmit the chunk, where $(M,n,\epsilon)$
represents a code over the DMC with $M$ messages, $n$ channel uses, and probability of error less than $\epsilon$.
When a sensor is requested to transmit, say, the $i$th chunk of the message, an honest sensor will use $\Gc_1$ to 
transmit the $i$th chunk.  A Byzantine sensor can choose to act honestly and use $\Gc_1$ to transmit the correct chunk, 
or it can transmit any other signal.

The second codebook $\Gc_2$ is a $(j,l,\epsilon)$ code used by the sensor in the verification
process. Specifically, to verify if a transmission represents correct information, the fusion center uses a random binning
technique. It distributes all possible chunks into $j$ bins and broadcasts
the bin index of each possible chunk to the sensors.  The fusion center then asks
$k$ sensors to transmit the bin index of the particular chunk that the fusion center is verifying.
An honest sensor will transmit the bin index to the fusion center using this second codebook $\Gc_2$.
  For fixed $j$, the code length $l$ is chosen sufficiently long for transmitting
the bin index accurately over the DMC.   A Byzantine sensor, if requested for the index,
again can transmit arbitrarily including acting honestly by using $\Gc_2$ to transmit the correct index.
The numbers $j$ and $k$ are functions of decoding error $\epsilon$
and are chosen sufficiently large
to ensure the fidelity of verification but not large enough to penalize the rate.  We comment on
the selection of them in Section~\ref{sub:events}.

The detailed transmission protocol is as follows.
\begin{enumerate}
\item[0)] The fusion center randomly selects a sensor to transmit the next chunk (starting at the first chunk).
\item If the selected sensor is honest, it transmits the entire chunk using the codebook $\Gc_1$. (If the selected
sensor is Byzantine, it can act arbitrarily).
\item The fusion center randomly places each element in the set of all possible chunks into one of $j$ bins.
The fusion center randomly selects $k$ sensors, and sends the binning to each of them. Each of those $k$ sensors then sends the bin index of the chunk back to the fusion center using code $\Gc_2$.
\item If more than half of the $k$ received bin indices match the bin index of the chunk that was received in step (1), the fusion center \emph{accepts} that chunk. Otherwise it \emph{declines} it.
\item If the chunk was accepted, the fusion center keeps the same sensor selected and moves on to the next chunk (go to step 1).
If it was declined. the fusion center randomly selects a new sensor and tries again with the same chunk (step 0).
\item Polling stops when all chunks have been received and accepted. To complete the coding process, the fusion center extracts the original message from the $v$ accepted chunks.
\end{enumerate}
Note that each time we run through steps (1) through (4), we use the channel $n/v+kl$ times.

In step (2), we have used a random binning procedure. This is different from the way
 such a procedure is often used, in which it is done just once during the construction of the code,
 but then the codebook is fixed. Here, we actually construct an entirely new random binning
 \emph{every time} we do step (2). This is necessary because if we used some fixed or
 deterministic binning, then if a Byzantine sensor is selected to transmit a chunk in step (1),
 it would know the binning to be used beforehand, so it could find a chunk in the
 same bin as the real chunk, which would make the verification useless. The probability that the
 Byzantine sensor selects a chunk different from the real chunk but in the same bin must be small,
 so we need {\em dynamic random binning}.

\subsection{Error Events and Error Probability Analysis}\label{sub:events}

We show next that, with appropriately chosen $n,v,j,l,k$ in the two codebooks, the
probability that a message is decoded incorrectly goes to zero, and the decoding process will
end with an average number of transmissions approximately $n+O(\epsilon n)$.  Thus with a message set of size $2^{nR}$, and
$R \ge C-\epsilon$, we have the proof of the main theorem.

To analyze the probability of error, we need to define some events. Events $\Amsc_1,\Amsc_2,\Amsc_3$ are the most basic ways in which errors can occur. $\Bmsc_1,\Bmsc_2,\Cmsc$ have to do with the conclusion the fusion center reaches, and thus determine how the coding will progress.
\begin{itemize}
\item $\Amsc_1$: A coding error occurs in step (1), \ie the transmitted chunk is different from the decoded one.
\item $\Amsc_2$: Of the $k$ bin indices that are decoded in step (2), less than half of them equal the bin index for the true chunk.
\item $\Amsc_3$: For a given pair of distinct chunks, they are both put into the same bin in step (2).
\item $\Bmsc_1$: The chunk is declined in step (3).
\item $\Bmsc_2$: A chunk is accepted in step (3) and that chunk is not the true one.
\item $\Cmsc$: The true chunk is transmitted in step (1).
\end{itemize}

The following lemma bounds the probabilities of
events relevant to the error analysis.

 \begin{lemma}\label{lemma:events} Define
\[{\setlength\arraycolsep{4pt}\begin{array}{lllll}
p_1&\defeq&\Pr(\Bmsc_1|\Cmsc),~~ p_2&\defeq&\Pr(\Bmsc_2|\Cmsc^c),\\
p_3&\defeq&\Pr(\Bmsc_2|\Cmsc).
\end{array}}\]
For sufficiently large $j$ and $k$, and no matter what the Byzantine sensors do, $\Pr(\Amsc_i)\le\epsilon$ for $i=1,2,3$, and
\[{\setlength\arraycolsep{3pt}\begin{array}{lllll}
p_1&\le&\Pr(\Amsc_1)+\Pr(\Amsc_2),~~
p_2 &\le& \Pr(\Amsc_2)+\Pr(\Amsc_3),\\
p_3&\le&\multicolumn{3}{l}{\Pr(\Amsc_1)(\Pr(\Amsc_2)+\Pr(\Amsc_3)).}
\end{array}}\]
\end{lemma}
\vspace{8pt}
\begin{proof}
Since $\Gc_1$ was constructed to have error probability less that $\epsilon$, $\Pr(\Amsc_1)\le\epsilon$.

Now we show $\Pr(\Amsc_2)\le\epsilon$ for sufficiently large $k$. Consider one of the $k$ sensors polled in step (2). It will be honest and a $\Gc_2$ error will not occur when it sends its bin index in step (2) with probability $(1-\beta)(1-\epsilon)$. These two events are sufficient (though not necessary) for the decoded bin index to be the real one. Therefore the probability that the decoded bin index is the real one is at least $(1-\beta)(1-\epsilon)$, so the probability that the decoded bin index is not the true one is no more than $\alpha\triangleq 1-(1-\beta)(1-\epsilon)$. Thus the number of decoded bin indices that are incorrect will be upper bounded by a binomial distribution with each one having probability $\alpha$ of being incorrect. Since $\beta<1/2$, for sufficiently small $\epsilon$, $\alpha<1/2$, so we will assume that this is the case. Thus
\begin{eqnarray}
\Pr(\Amsc_2)&\le&\sum_{i=k/2}^k\binom{k}{i}\alpha^i(1-\alpha)^{k-i}\nonumber\\
&\le&\binom{k}{k/2}(1-\alpha)^k\sum_{i=k/2}^k\left(\frac{\alpha}{1-\alpha}\right)^i\label{eq:pra2}\\
&=&\frac{1-\alpha}{1-2\alpha}\binom{k}{k/2}\left(\alpha^{k/2}(1-\alpha)^{k/2}-\alpha^{k+1}\right)\nonumber\\
&\le&\frac{1-\alpha}{1-2\alpha}4^{k/2}\alpha^{k/2}(1-\alpha)^{k/2}\label{eq:pra4}
\end{eqnarray}
where \eqref{eq:pra2} holds because $\binom{k}{i}\le\binom{k}{k/2}$ for all $i\in\{0,\cdots,k\}$, and \eqref{eq:pra4} holds because $\alpha<1/2$, so the denominator $1-2\alpha$ is positive, so the $-\alpha^{k+1}$ term can be dropped, and because $\binom{2m}{m}\le 4^m$ for all $m$. Thus if
\[k\ge 2\frac{\log\big(\frac{1-2\alpha}{1-\alpha}\ \epsilon\big)}{\log(4\alpha(1-\alpha))},\]
then $\Pr(\Amsc_2)\le\epsilon$.

Next we show $\Pr(\Amsc_3)\le\epsilon$ for sufficiently large $j$. Since there are $j$ bins, the probability that two different chunks are put into the same bin in step (2) is $1/j$. Thus if $j\ge 1/\epsilon$, $\Pr(\Amsc_3)\le\epsilon$.

Note that $p_1$ is the probability that the received chunk is declined in step (3) given the true chunk was transmitted in step (1). One way for this to happen is for there to be a coding error in step (1), \ie $\Amsc_1$ occurs, so the received chunk will not be the true chunk, so the polled sensors may not confirm it. Note that a coding error does not necessitate the chunk being declined, but it does cover a large set of the ways it could happen. If $\Amsc_1$ does not occur, then the received chunk is the true one, so the chunk could only be declined if the majority of the bin indices received in step (2) do not match the true chunk, \ie $\Amsc_2$ occurs. Thus
\[p_1\le\Pr(\Amsc_1\cup\Amsc_2)\le\Pr(\Amsc_1)+\Pr(\Amsc_2).\]

Next, $p_2$ is the probability that an incorrect chunk is accepted given that an incorrect chunk is transmitted in step (1). If more than half of the decoded bin indices are incorrect ($\Amsc_2$), then those incorrect bin indices might confirm the incorrect chunk. If not, then the only way for the incorrect chunk to be accepted is for it to fall into the same bin as the true chunk ($\Amsc_3$). Thus
\[p_2\le\Pr(\Amsc_2\cup\Amsc_3)\le\Pr(\Amsc_2)+\Pr(\Amsc_3).\]

Finally, $p_3$ is the probability that an incorrect chunk is accepted given that the correct chunk is transmitted in step (1). In order for this to happen, the decoded chunk must not be the true one, so a coding error must occur ($\Amsc_1$). In addition, for that decoded incorrect chunk to be accepted, more than half of the decoded bin indices must be incorrect ($\Amsc_2$) or the incorrect chunk must fall into the same category as the real one ($\Amsc_3$). Thus
\[p_3\le\Pr(\Amsc_1)\Pr(\Amsc_2\cup\Amsc_3)\le\Pr(\Amsc_1)(\Pr(\Amsc_2)+\Pr(\Amsc_3)).\]
\end{proof}

\begin{figure}
\centerline{
\begin{psfrags}
\psfrag{pb}[c]{$\scriptstyle{p_1\beta}$}
\psfrag{pob}[c]{$\scriptstyle{p_1(1-\beta)}$}
\psfrag{oppb}[c]{$\scriptstyle{(1-p_2)\beta}$}
\psfrag{oppob}[c]{$\scriptstyle{(1-p_2)(1-\beta)}$}
\psfrag{pp}[c]{$\scriptstyle{p_2}$}
\psfrag{ppp}[c]{$\scriptstyle{p_3}$}
\psfrag{opppp}[c]{$\scriptstyle{1-p_1-p_3}$}
\psfrag{oo}[c]{$\scriptstyle{1}$}
\psfrag{aa}[b][B][1.3]{$0$}
\psfrag{aap}[b][B][1.3]{$0'$}
\psfrag{bb}[b][B][1.3]{$1$}
\psfrag{bbp}[b][B][1.3]{$1'$}
\psfrag{cc}[b][B][1.3]{$2$}
\psfrag{ccp}[b][B][1.3]{$2'$}
\psfrag{vv}[b][B][1.3]{$v$}
\psfrag{vvp}[b][B][1.3]{$v'$}
\psfrag{ee}[b][B][1.3]{$e$}
\scalefig{0.45}\epsfbox{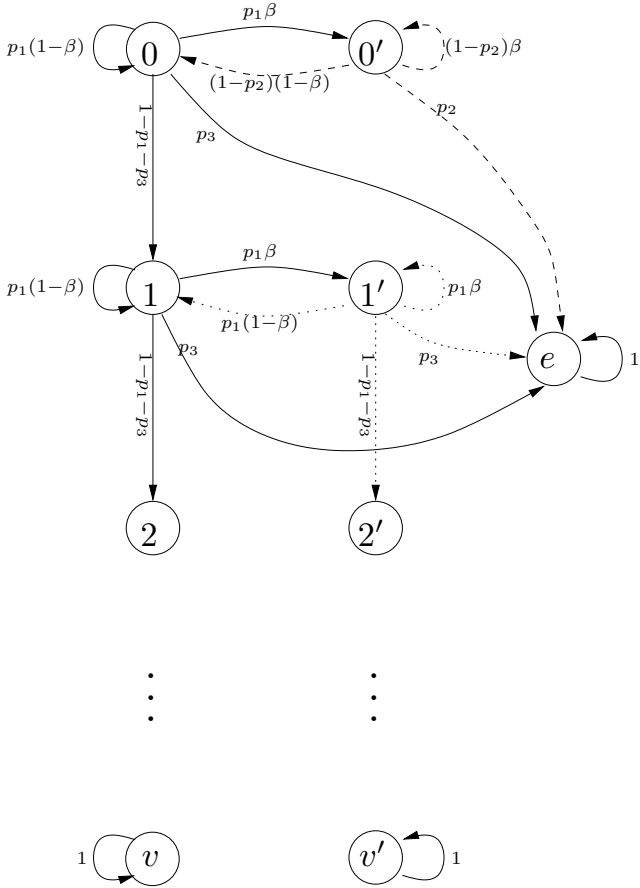}
\end{psfrags}}
\caption{The Markov decision process used to find the error probability. Dashed lines from a state represent the Byzantine sensor choosing to send erroneous information, and dotted lines represent the Byzantine sensor choosing to send true information.}
\label{fig:markov}
\end{figure}

As the coding scheme commences, it moves through a number of different states, depending on the number chunks the fusion center has received thus far, and whether the selected sensor is Byzantine. Depending on the exact sequence of events, the fusion center might remain at a certain state for some time, requesting the same chunk several times until it finds an honest sensor. The progress is probabilistic because every time the fusion center selects a sensor it might be Byzantine or honest, and every time it receives a transmission, a transmission error might or might not occur. In fact, the progress of the coding scheme can be modeled as a Markov process. In particular, it will be a Markov decision process, because a Byzantine sensor, if it is selected to transmit a chunk, has some choice about what to transmit. That choice will influence the probabilities of future events. The Markov decision process that we will use to analyze the error probability of this scheme is diagrammed in Fig. \ref{fig:markov}.

The process will have $2v+3$ states. State $i$, for $i=0,\cdots,v$ represents the fusion center having successfully received $i$ true chunks and the currently selected sensor is honest. State $i'$ is the same except the currently selected sensor is Byzantine. Finally, state $e$ represents the fusion center having accepted at least one false chunk. The decision for the Markov decision process will be whether a Byzantine sensor, if it is asked to send a chunk in step (1), chooses to send the true chunk or not. Thus a decision will only be made when a Byzantine sensor has been selected, \ie we are in one of the $i'$ states.

States $v$, $v'$, and $e$ will be terminal states, so an error will occur if we reach state $e$ before state $v$ or $v'$.
Define
\begin{eqnarray*}
e_i &\defeq& \Pr(\mbox{error occurs starting from state $i$}),\\
e_i' &\defeq& \Pr(\mbox{error occurs starting from state $i'$}).
\end{eqnarray*}
In executing the Markov decision process, the Byzantine sensors make decisions to maximize the probability of error. At the very beginning of the coding scheme, we select a sensor which will be with probability $1-\beta$ honest and probability $\beta$ Byzantine. Thus, the total probability of error is
\[
P_e=(1-\beta)e_0+\beta e_0'.
\]

From state $i$, with probability $p_1$ the chunk will be declined. The fusion center then selects a new sensor, which will be Byzantine with probability $\beta$ and honest with probability $1-\beta$. Thus we transition to state $i'$ with probability $p_1\beta$ and back to state $i$ with probability $p_1(1-\beta)$. With probability $p_3$, an incorrect chunk is accepted, so we transition to state $e$. Finally, with probability $1-p_1-p_3$ the true chunk is accepted, so we transition to state $i+1$.

From state $i'$, the transition probabilities depend on the decision. If the Byzantine sensor chooses not to send the true chunk, then with probability $p_2$ the false chunk will be accepted, so we transition to state $e$. Otherwise, the fusion center selects a new sensor. Thus with probability $(1-p_2)\beta$ we return to state $i'$, and with probability $(1-p_2)(1-\beta)$ we transition to state $i$. If the Byzantine sensor decides to send the true chunk, then the transition probabilities are essentially the same as they were from state $i$; with probability $p_1\beta$ we return to state $i'$, with probability $p_1(1-\beta)$ we transition to state $i$, with probability $p_3$ we transition to state $e$, and with probability $1-p_1-p_3$ we transition to state $i+1'$.

From these transition probabilities, we see that
\[
{\setlength\arraycolsep{0pt}\begin{array}{llllll}
e_i&\ =\ &\multicolumn{4}{l}{p_3+(1-p_1-p_3)e_{i+1}+p_1\beta e_i'+p_1(1-\beta)e_i,}\\
e_i'&\ =\ &\max\{&p_2&\:+\:&(1-p_2)\beta e_i'+(1-p_2)(1-\beta)e_i,\\
&&&p_3&\:+\:&(1-p_1-p_3)e_{i+1}'+p_1\beta e_i'\\
&&&&\:+\:&p_1(1-\beta)e_i\}.
\end{array}}\]
The maximum represents the Byzantine sensors always making the decision that maximizes the error probabilities. In addition, if we arrive at either state $v$ or $v'$, the fusion center has received the entire message without error, so $e_v=e_v'=0$.

\subsection{Code Rate}\label{sub:rate}

We also need to consider the rate of this code. To show that the rate can be made arbitrarily close to $C$, we need to show that the expected number of channel uses $\mbbE(N)$ converges to $n$ as $\epsilon$ goes to zero. Each time a chunk is transmitted (\ie each time we run through steps (1) to (4)), the channel is used $n/v+kl$ times. All we need to know is the expected number of chunks that are transmitted in the entire coding scheme. To find this, we will use a similar Markov decision process as the one described above. The only differences lie in the fact that we are not interested in whether an error occurs, only in how long it takes to finish. Thus we remove state $e$ and redefine states $i$ and $i'$ to represent the fusion center having accepted $i$ states, but with all of them not necessarily correct. Thus every time we would transition to state $e$, we actually transition somewhere else. For instance, if we are in state $i'$ and the Byzantine sensors choose to send erroneous information, then with probability $p_2$, the chunk is accepted, so we transition to state $i+1'$ instead of $e$. Let $q_i$ and $q_i'$ be the expected number of steps made in the Markov decision process before reaching one of the terminal states ($v$ or $v'$) given that we start at state $i$ or $i'$ respectively and the Byzantine sensors make decisions that maximize the expected number of steps. Then
\beq\label{eq:qidef}{\setlength\arraycolsep{0pt}\begin{array}{llllll}
q_i&\ =\ &\multicolumn{4}{l}{1+(1-p_1)q_{i+1}+p_1\beta q_i'+p_1(1-\beta)q_i,}\\
q_i'&\ =\ &\max\{&1&\:+\:&p_2q_{i+1}'+(1-p_2)\beta q_i'\\
&&&&\:+\:&(1-p_2)(1-\beta)q_i,\\
&&&1&\:+\:&(1-p_1)q_{i+1}'+p_1\beta q_i'+p_1(1-\beta)q_i\}.
\end{array}}\eeq
Again, $q_v=q_v'=0$.

\begin{lemma}[Average Code Length] There exist $n$, $v$, $j$, and $k$ as functions of $\epsilon$ such that the error probability $P_e\to 0$ and the expected number of channel uses $\mbbE(N)\to n$ as $\epsilon\to 0$.
\end{lemma}
\begin{proof}
Take $j$ and $k$ large enough for Lemma~\ref{lemma:events} to hold, and $n$ and $v$ such that
\beq\label{eq:vn}
\frac{2}{\epsilon}\ge v \ge\frac{1}{\epsilon},~~n\ge\frac{klv}{\epsilon}.
\eeq

We define $f_i,f_i'$ for $i=0,\cdots,v$ as follows.  Let $f_v\triangleq f_v'\triangleq 0$ and for $i<v$,
{\setlength\arraycolsep{2pt}\begin{eqnarray}
f_i&\triangleq&p_3+(1-p_1-p_3)f_{i+1}+p_1\beta f_i'+p_1(1-\beta)f_i,\quad\label{eq:fidef}\\
f_{i,a}'&\triangleq&p_2+\beta f_i'+(1-\beta)f_i,\label{eq:fiadef}\\
f_{i,b}'&\triangleq&p_3+(1-p_1-p_3)f_{i+1}'+p_1\beta f_i',\label{eq:fibdef}\\
f_i'&\triangleq&\max\{f_{i,a}',f_{i,b}'\label{eq:fipdef}\}.
\end{eqnarray}
The only difference between $f_i,f_i'$ and $e_i,e_i'$ is that the $(1-p_2)$ factors have been dropped from the second two terms in \eqref{eq:fiadef}. Thus $e_i\le f_i,e_i'\le f_i',$ for all i. Fix some \hbox{$i\in\{0,\cdots,v-1\}$.} If $f_i'=f_{i,a}'$, then by \eqref{eq:fiadef}}
\beq\label{eq:fipfi}
f_i'=\frac{p_2}{1-\beta}+f_i.
\eeq
Combining this with \eqref{eq:fidef} gives
\beq\label{eq:fifi1}
f_i=f_{i+1}+\frac{p_3}{1-p_1}+\frac{p_1p_2\beta}{(1-p_1)(1-\beta)},
\eeq
which with \eqref{eq:fipfi} produces
\begin{eqnarray}
f_i'&=&f_{i+1}+\frac{p_3}{1-p_1}+\frac{p_1p_2\beta}{(1-p_1)(1-\beta)}+\frac{p_2}{1-\beta}\nonumber\\
&=&f_{i+1}+\frac{p_3}{1-p_1}+\frac{p_2(1-p_1(1-\beta))}{(1-p_1)(1-\beta)}.\label{eq:fipa}
\end{eqnarray}
If $f_i'=f_{i,b}$, then combining \eqref{eq:fidef} with \eqref{eq:fibdef} gives
\beq\label{eq:fipb}
f_i'=\frac{p_3}{1-p_1}+p_1(1-\beta)f_{i+1}+(1-p_1(1-\beta))f_{i+1}'.
\eeq
Note that \eqref{eq:fipa} and \eqref{eq:fipb} are what $f_i'$ would be if $f_i'$ equaled $f_{i,a}'$ or $f_{i,b}'$ respectively. However, these expressions are not necessarily equal to $f_{i,a}'$ and $f_{i,b}'$, because we have used \eqref{eq:fidef} to derive both of them, which contains the real value of $f_i'$. Still, because of the definition of $f_i'$ in \eqref{eq:fipdef}, the larger of \eqref{eq:fipa} and \eqref{eq:fipb} will be the true value of $f_i'$.

We will now show by induction that $f_i'=f_{i,a}'$ for $i=0,\cdots,v-1$. For $i=v-1$, since $f_v=f_v'=0$, it is clear that the expression in \eqref{eq:fipa} is larger than that in \eqref{eq:fipb}, so $f_{v-1}'=f_{v-1,a}'$. Now we assume that $f_{i+1}'=f_{i+1,a}'$ and show that $f_i'=f_{i,a}'$. By \eqref{eq:fipfi},
\[f_{i+1}'=\frac{p_2}{1-\beta}+f_{i+1}.\]
Thus, if $f_i'=f_{i,b}'$, \eqref{eq:fipb} becomes
\begin{eqnarray*}
f_i'&=&\frac{p_3}{1-p_1}+p_1(1-\beta)f_{i+1}\\
&&+(1-p_1(1-\beta))\left(\frac{p_2}{1-\beta}+f_{i+1}\right)\\
&=&\frac{p_3}{1-p_1}+f_{i+1}+\frac{p_2(1-p_1(1-\beta))}{1-\beta}.
\end{eqnarray*}
Since the expression in \eqref{eq:fipa} is larger than this, $f_i'=f_{i,a}'$.

Therefore \eqref{eq:fifi1} holds for $i=0,\cdots,v-1$, so 
\beq\label{eq:fiv}
f_i=\left(\frac{p_3}{1-p_1}+\frac{p_1p_2\beta}{(1-p_1)(1-\beta)}\right)(v-i).\eeq
Thus
\begin{eqnarray}
P_e&=&(1-\beta)e_0+\beta e_0'\nonumber\\
&\le&(1-\beta)f_0+\beta f_0'\nonumber\\
&=&f_0+\frac{p_2\beta}{1-\beta}\label{eq:pe3}\\
&=&\frac{p_1p_2\beta+p_3(1-\beta)}{(1-p_1)(1-\beta)}\ v+\frac{p_2\beta}{1-\beta}\label{eq:pe4}\\
&\le&\frac{4\epsilon^2\beta+2\epsilon^2(1-\beta)}{(1-2\epsilon)(1-\beta)}\left(\frac{2}{\epsilon}\right)+\frac{2\epsilon\beta}{1-\beta}\label{eq:pe5}\\
&=&\left(\frac{8\beta+4(1-\beta)}{(1-2\epsilon)(1-\beta)}+\frac{2\beta}{1-\beta}\right)\epsilon\nonumber
\end{eqnarray}
where \eqref{eq:pe3} is from \eqref{eq:fipfi}, \eqref{eq:pe4} is from \eqref{eq:fiv}, and \eqref{eq:pe5} is from Lemma~\ref{lemma:events} and \eqref{eq:vn}. Thus $P_e\to 0$ as $\epsilon\to 0$.

Now we analyze $q_i,q_i'$ to find $\mbbE(N)$. Combining the expression for $q_i$ in \eqref{eq:qidef} with either expression for $q_i'$ in the maximum in \eqref{eq:qidef} yields expressions of the form
\begin{eqnarray}
q_i&=&1+\gamma+\delta q_{i+1}+(1-\delta)q_{i+1}',\label{eq:qigam}\\
q_i'&=&1+\gamma'+\delta' q_{i+1}+(1-\delta')q_{i+1}',\label{eq:qipgam}
\end{eqnarray}
where $\gamma,\gamma'\ge 0$ and $\delta,\delta'\in[0,1]$. The quantity $\gamma$ represents the expected number of state transitions between states $i$ and $i'$ before moving on to state $i+1$ or $i+1'$, given that we start at state $i'$, and $\delta$ represents the probability that when we do transition away from states $i$ and $i'$, we go to state $i+1$ and not $i+1'$. The quantities $\gamma'$ and $\delta'$ are the same except starting at state $i'$. Obviously, the values of these will depend on which element of the maximum is larger, but for our current purposes it only matters that the expressions will have this form.

We will now show by induction that $q_i-q_{i+1}\ge 1$ and $q_i'-q_{i+1}'\ge 1$ for $i=0,\cdots,v-1$. First consider $i=v-1$. $q_v=q_v'=0$, so by \eqref{eq:qigam} and \eqref{eq:qipgam}, $q_{v-1}=1+\gamma$ and $q_{v-1}'=1+\gamma'$. Thus $q_{v-1}-q_v\ge 1$ and $q_{v-1}'-q_v'\ge 1$. Now we assume that $q_{i+1}-q_{i+2}\ge 1$ and $q_{i+1}'-q_{i+2}'\ge 1$ and show that $q_i-q_{i+1}\ge 1$ and $q_i'-q_{i+1}'\ge 1$. By assumption and \eqref{eq:qigam},
\begin{eqnarray*}
q_i-q_{i+1}&=&\delta(q_{i+1}-q_{i+2})+(1-\delta)(q_{i+1}'-q_{i+2}')\\
&\ge&\delta+(1-\delta)\\
&=&1.
\end{eqnarray*}
Similarly by \eqref{eq:qipgam},
\begin{eqnarray*}
q_i'-q_{i+1}'&=&\delta'(q_{i+1}-q_{i+2})+(1-\delta')(q_{i+1}'-q_{i+2}')\\
&\ge&\delta'+(1-\delta')\\
&=&1.
\end{eqnarray*}
Thus $q_i-q_{i+1}\ge 1$ and $q_i'-q_{i+1}'\ge 1$ for $i=0,\cdots,v-1$. In particular, $q_{i+1}'\le q_i'-1$.

Suppose the first element of the maximum is larger in \eqref{eq:qidef}. Then
\begin{eqnarray*}
q_i'&=&1+p_2q_{i+1}'+(1-p_2)\beta q_i'+(1-p_2)(1-\beta)q_i\\
&\le&1+p_2(q_i'-1)+(1-p_2)\beta q_i'+(1-p_2)(1-\beta)q_i.
\end{eqnarray*}
This can be rewritten
\beq\label{eq:qipqi}
q_i'\le\frac{1}{1-\beta}+q_i.
\eeq
Now suppose the second element of the maximum is larger in \eqref{eq:qidef}. Then
\begin{eqnarray*}
q_i'&=&1+(1-p_1)q_{i+1}'+p_1\beta q_i'+p_1(1-\beta)q_i\\
&\le&1+(1-p_1)(q_i'-1)+p_1\beta q_i'+p_1(1-\beta)q_i.
\end{eqnarray*}
This can also be rewritten to \eqref{eq:qipqi}, so \eqref{eq:qipqi} must hold no matter which value is larger in the maximum in \eqref{eq:qidef}. Thus
\[q_i\le 1+(1-p_1)q_{i+1}+p_1\beta \left(\frac{1}{1-\beta}+q_i\right)+p_1(1-\beta)q_i.\]
This can be rewritten
\[q_i\le 1+\frac{p_1}{(1-p_1)(1-\beta)}+q_{i+1},\]
so
\beq\label{eq:qiv}
q_i\le \left(1+\frac{p_1}{(1-p_1)(1-\beta)}\right)(v-i).\eeq
Let $V$ be the random variable denoting the total number of chunks that are requested in the entire coding session. Since we start at state 0 with probability $1-\beta$ and at state $0'$ with probability $\beta$,
{\setlength{\arraycolsep}{2pt}\begin{eqnarray}
\mbbE(V)&=&(1-\beta)q_0+\beta q_0'\nonumber\\
&\le&q_0+\frac{\beta}{1-\beta}\label{eq:en2}\\
&\le&\left(1+\frac{p_1}{(1-p_1)(1-\beta)}\right)v+\frac{\beta}{1-\beta}\label{eq:en3}
\end{eqnarray}
where \eqref{eq:en2} is from \eqref{eq:qipqi} and \eqref{eq:en3} is from \eqref{eq:qiv}. Thus
\begin{eqnarray}
\mbbE(N)&=&\mbbE(V)\left(\frac{n}{v}+kl\right)\nonumber\\
&\le&\left[\left(1+\frac{p_1}{(1-p_1)(1-\beta)}\right)v+\frac{\beta}{1-\beta}\right]\left(\frac{n}{v}+kl\right)\nonumber\\
&=&n\left[1+\frac{p_1}{(1-p_1)(1-\beta)}+\frac{\beta}{1-\beta}\ \frac{1}{v}\right.\nonumber\\
&&\quad\left.+\left(1+\frac{p_1}{(1-p_1)(1-\beta)}\right)\frac{klv}{n}+\frac{\beta}{1-\beta}\ \frac{kl}{n}\right]\nonumber\\
&\le&n\left[1+\frac{2\epsilon}{(1-2\epsilon)(1-\beta)}+\frac{\beta}{1-\beta}\ \epsilon\right.\nonumber\\
&&\quad\left.+\left(1+\frac{2\epsilon}{(1-2\epsilon)(1-\beta)}\right)\epsilon+\frac{\beta}{1-\beta}\ \epsilon^2\right]\label{eq:en5}\\
&=&n\left[1+\left(\frac{2(1+\epsilon)}{(1-2\epsilon)(1-\beta)}+\frac{\beta(1+\epsilon)}{1-\beta}+1\right)\epsilon\right]\nonumber
\end{eqnarray}}
where \eqref{eq:en5} is from Lemma~\ref{lemma:events} and \eqref{eq:vn}. Thus $\mbbE(N)\to n$ as $\epsilon\to 0$.
\end{proof}

Therefore the rate of this code,
\[\frac{nR}{\mbbE(N)},\]
converges to $R$ as $\epsilon$ goes to 0. Thus $C$ is achievable.

\section{Conclusion}\label{sec:conc}
We showed in this paper that, by cooperative sensor fusion, the presence of
Byzantine sensors can be completely mitigated when the Byzantine sensor population
is less than half of the total number of sensors, but no information can be 
transmitted when at least half of the sensors are Byzantine.  We proposed a ``transmit-then-verify''
scheme that forces a Byzantine sensor to either act honestly or reveal its Byzantine
identity.  The key of this idea is the use of random binning in sensor polling.
Note that the random binning in our strategy is not a random coding argument;
it is an actual randomized transmission protocol.

Several simple generalizations can be made. The network does not
have to contain infinite number of sensors.  For a finite size network,
we will assume that a deterministic  $\beta$ fraction of the sensors are Byzantine.
In that case, all the sensors can be polled when verifying a transmission. Thus if less than 
half of the sensors are Byzantine, information will always be correctly verified. This requires a 
constant and hence asymptotically negligible number of channel uses, so polling every sensor instead of a random subset does 
not effect the rate.
We can also relax the assumption that the consensus is perfect by assuming that
there is a fraction of sensors that are are misinformed as in \cite{Yang&Tong:05IT}. In such a circumstance, a similar 
coding algorithm as the one described in this paper can be used, and the full channel capacity can be achieved 
as long as the correctly informed honest sensors outnumber the Byzantine sensors, though
the proof of this is nontrivial.

\bibliographystyle{ieeetr}
{
\bibliographystyle{ieeetr}
\bibliography{\ACSP/Reference/Bibs/Journal,\ACSP/Reference/Bibs/Conf,%
\ACSP/Reference/Bibs/Book,\ACSP/Reference/Bibs/Misc,\ACSP/Reference/Bibs/ACSP-J,\ACSP/Reference/Bibs/ACSP-C}
}

\end{document}